 \documentclass[final,3p,times,twocolumn]{elsarticle}
\usepackage{graphicx}
\usepackage{amssymb}
\usepackage{siunitx}
\usepackage{xcolor}
\usepackage[normalem]{ulem}
\journal{Ultramicroscopy}

\begin{document}

\begin{frontmatter}


\title{Quantifying the Performance of a Hybrid Pixel Detector with GaAs:Cr Sensor for Transmission Electron Microscopy}



\author[1]{Kirsty A. Paton\corref{cor1}}
\cortext[cor1]{Corresponding author}
\ead{kirsty.paton@glasgow.ac.uk}
\author[2]{Mathew C. Veale}
\author[3]{Xiaoke Mu}
\author[4,5]{Christopher S. Allen}
\author[1]{Dzmitry Maneuski}
\author[3,6]{Christian K{\"u}bel}
\author[1]{Val O'Shea}
\author[4,5]{Angus I. Kirkland}
\author[1]{Damien McGrouther}

\address[1]{Scottish Universities Physics Alliance, School of Physics and Astronomy, University of Glasgow, G12 8QQ, UK}
\address[2]{UKRI Science \& Technology Facilities Council, Rutherford Appleton Laboratory, Didcot, OX11 0QX, UK}
\address[3]{Karlsruhe Nano Micro Facility (KNMF), Karlsruhe Institute of Technology, Hermann-von-Helmholtz-Platz 1, 76344, Eggenstein-Leopoldshafen, Germany}
\address[4]{Department of Materials, University of Oxford, Parks Road, Oxford, OX1 3PH, UK}
\address[5]{Electron Physical Sciences Imaging Centre, Diamond Lightsource Ltd., Didcot, OX11 0DE, UK}
\address[6]{Department of Materials and Earth Science, Technische Universität Darmstadt and Karlsruhe Institute of Technology, Otto-Berndt-Str. 3, 64287 Darmstadt, Germany}

\begin{abstract}
 Hybrid pixel detectors (HPDs) have been shown to be highly effective for diffraction-based and time-resolved studies in transmission electron microscopy, but their performance is limited by the fact that high-energy electrons scatter over long distances in their thick Si sensors. 
 An advantage of HPDs compared to monolithic active pixel sensors (MAPS) is that their sensor does not need to be fabricated from Si. 
 We have compared the performance of the Medipix3 HPD with a Si sensor and with a GaAs:Cr sensor using primary electrons in the energy range of 60~-~300keV. 
 We describe the measurement and calculation of the detectors' modulation transfer function (MTF) and detective quantum efficiency (DQE), which show that the performance of the GaAs:Cr device is markedly superior to that of the Si device for high-energy electrons. 
% Analysis of the detectors' response to individual electrons provides insight into the interactions of electrons with thick sensors and how these determines the performance of HPDs.
\end{abstract}


\end{frontmatter}


\section{Introduction}
\label{S:1}

The development of direct electron detectors (DEDs) over the past twenty years has opened up new experimental possibilities in electron microscopy, leading to significant advances in various fields \cite{Merk2016, Jiang2018}. 
Key to this success is increased sensitivity to incident electrons, which facilitates electron counting, compared to indirect scintillator-coupled detectors. 
This is invaluable when electron dose to the sample is low ($\leq~10\,e/\AA{}^{2}$) and also highly advantageous when performing quantitative analysis \cite{Mcmullan2009a, Song2019, Chen2016}.
%Furthermore, the acquisition and processing of data is more flexible and convenient than is the case with film, which was previously preferred where high spatial resolution was critical \cite{Mcmullan2009, Faruqi2001}.
DEDs can be broadly divided into two categories: hybrid pixel detectors (HPDs) and monolithic active pixel sensors (MAPS).
The latter have had great impact, substantially improving the resolution limit of cryogenic electron microscopy (cryoEM) at higher ($\geq$~200kV) accelerating voltages \cite{Bai2015, Li2013}. 

However, the former are more suitable for a wider range of applications, particularly those requiring high frame-rates, a linear response to high electron flux and radiation hardness.
HPDs consist of an application specific integrated circuit (ASIC), which contains the signal-processing and readout electronics, bump-bonded to a thick ($\geq$~\SI{300}{\micro\meter}) sensor that protects the ASIC from the incident electrons, making them highly radiation resilient.
On-pixel signal processing circuitry makes them capable of high (typically kHz) frame-rates and electron counting at MHz rates \cite{Frojdh2014, Tate2016}.
This makes them highly effective sensors for capturing fast (1ms) dynamics in a conventional transmission electron microscope (TEM) \cite{Kirkland2019}, and they show the potential to record processes at timescales of $\leq\SI{1}{\micro\second}$ \cite{Beacham2011, Paterson2020}.
Their ability to maintain a linear response even when subjected to high ($\geq$~1000$~e\textrm{/pixel/s}$) electron flux means they are suitable for use in a variety of diffraction-based experiments \cite{Tate2016}. 
They have been successfully used for micro-electron diffraction (microED) in structural biology \cite{Tinti2018} and 4D scanning transmission electron microscopy (4D-STEM), in both convergent and nano beam electron diffraction modes \cite{Fang2019, Temple2018, Krajnak2016} in materials science.
Beyond this, they have facilitated the application of 4D-STEM to biological samples \cite{Bucker2020, Zhou2020} and are promising for use in electron energy-loss spectroscopy \cite{Plotkin-Swing2020}.

Neither type of DED is capable of maximum performance across the full range of incident electron energies available on current generation TEM instruments (60~-~300keV). 
MAPS devices consist of a thin Si sensor, $\leq$~\SI{50}{\micro\meter} in thickness, and minimal on-pixel electronics, which means they have small, usually $<$\SI{15}{\micro\meter} in pitch, pixels \cite{Faruqi2005a}.
High-energy ($\geq$~200keV) electrons are transmitted through their thin sensors with minimal backscatter and there is little lateral spread in the signal produced by the small amount of energy that they deposit in the sensor \cite{Mcmullan2009b, Kuijper2015}. 
These factors mean they offer excellent imaging performance for high-energy electrons, making them the detector of choice for high-energy cryoEM. 
At lower ($\leq$120keV) electron energies, their performance deteriorates due to increased lateral scatter in and backscatter from the sensor, making them less suitable for low-energy cryoEM \cite{Peet2019, Naydenova2019} and studies of materials sensitive to sputtering damage \cite{Egerton2004, Cheng2020}.
Furthermore, the simplicity of the on-pixel electronics limits frame rates and, as they count electrons by identifying pixel clusters in sparsely populated frames, their count rates.
Current monolithic devices used in electron microscopy typically have sub-kHz frame rates, although the recently developed 4D Camera has a maximum frame rate of of 87kHz \cite{Ciston2019}. 
This has the drawback of producing large volumes of empty data that are computationally expensive to manage and process, whereas newer HPDs are able to operate in a data-driven mode whereby only those pixels that record a hit are read out, reducing the size of the datasets that are produced \cite{Poikela2014}.
The maximum electron fluence to which monolithic device can be exposed is also limited by the fragility of their on-pixel electronics \cite{Mcmullan2009a, Gao2019, Gallagher-Jones2019}.

HPDs, such as Medipix3 \cite{Ballabriga2013}, have been shown to match and even surpass the performance of an ideal detector when used with low-energy electrons \cite{Mir2017}.
However, for sensors sufficiently thick to protect the ASIC, high-energy electrons travel long distances and are counted by multiple pixels, causing a degradation in performance \cite{Tinti2018, McMullan2007}.
The lateral scattering and penetration depth of an incident electron is inversely proportional to the average atomic number (Z) of the sensor.
Unlike monolithic devices, HPDs can have sensors made of materials other than Si (Z~=~14).
HPDs with high-Z sensors should be capable of improved imaging performance for incident electrons across a wider range of energies, as the spatial distribution of the signal produced by high-energy electrons would be more localised \cite{Mcmullan2008}.
This would increase the versatility of HPDs and, combined with their advantages relative to MAPS detectors, they would have the potential to be ``universal'' detectors for transmission electron microscopy, suitable for almost all applications at all accelerating voltages.

Increasing the Z of the sensor may also have a negative impact on performance, through decreased efficiency due to increased backscatter \cite{Mcmullan2009}.
In this article, we investigate the extent to which the performance of HPDs can be improved by using a high-Z, specifically a GaAs:Cr (average Z~=~32), sensor. 
We begin by describing in detail procedures suitable for performing measurements of the modulation transfer function (MTF) and detective quantum efficiency (DQE) of HPDs, for the purpose of characterising their imaging performance. 
We then compare the imaging performance of \SI{500}{\micro\meter} thick GaAs:Cr and Si sensors bonded to Medipix3 ASICs for electrons at energies of 60~-~300keV.
Finally, we offer a comparison of their performance under uniform illumination and discuss some of the challenges associated with the use of high-Z sensors for imaging applications.

%In addition to count-rate, frame-rate and robustness, a key consideration guiding choice of detector is the energy of the electrons with which the detector is to be used. 
%Neither HPDs nor MAPS devices perform well across all electron energies that are typically used in a TEM.
%, as well as analysis of pixel clusters due to individual electrons recorded by the two detectors. As part of the present work, we outline suitable approaches to measuring the MTF, noise power spectrum (NPS) and DQE of HPDs.
%Although the performance of the Medipix3 bonded to a \SI{300}{\micro\meter} Si sensor has been previously characterised with low-energy electrons \cite{Mir2017}, we have included low-energy results here for the \SI{500}{\micro\meter} Si device, as, unlike a \SI{500}{\micro\meter} Si sensor, a \SI{300}{\micro\meter} sensor is insufficiently thick to protect the ASIC from 300keV electrons \cite{McMullan2007}. 
%The previously reported results cannot be fairly compared with the results for the GaAs:Cr device presented here, which is of interest for its potential to perform well across a range of electron energies, up to and including 300keV.

\section{Detector Structure}
\label{S:2}

The Medipix3RX ASIC (henceforth referred to as Medipix3) consists of an array of 256~$\times$~256 \SI{55}{\micro\meter} pitch pixels \cite{Ballabriga2013}.
The signal-processing circuitry present on each pixel has an analogue front-end and a digital back-end.
In the analogue section, the charge induced in a pixel due to an incident electron is amplified and converted into a shaped voltage pulse.
When the detector is operating in single-pixel mode (SPM), this is registered as a hit if it surpasses a user set threshold, and one of the Linear Feedback Shift Registers in the digital back-end is incremented.
During readout of a frame, the register acts as a shift register to readout the number of hits it has recorded during data acquisition.
Used with the Merlin readout system \cite{Plackett2013}, the detector is capable of frame greater than 1~kHz depending on counter bit-depth, with on-pixel count rates being determined by the ASIC settings \cite{Frojdh2014}.  
The digital back-end contains two registers and up to two thresholds, TH0 and TH1 can be set.
%In principal, one can set TH0 just above the level of the detector' noise and TH1 at an arbitrary higher value (e.g. half the primary electron energy) permitting a course measurement of the energy deposited on a pixel.
Alternatively, a single threshold can be used, with the two registers working in tandem such that while one is acting as a counter the other is operating as a shift register, permitting continuous acquisition of data with no dead time.
%In another variation of SPM, the two registers, which can otherwise be configured as 2~$\times$~1-bit, 2~$\times$~6-bit or 2~$\times$~12-bit counters, can be combined to act as a single 1~$\times$~24-bit counter.

The detector's other main mode of operation, is a charge summing mode (CSM), where neighbouring pixels pool their individual signal-processing circuitry and attempt to allocate incident electrons to a single pixel.
Each pixel compares the voltage pulse produced in its analogue front end to TH0 but also sends copies of this pulse to summing nodes that are effectively located at its corners.
At each node, the voltage pulses produced by the four pixels that share a corner are summed.
If the summed voltage pulses surpass TH1, then the counter of the pixel identified has having the largest share of the deposited energy is incremented.
The pixel which registers the most energy is identified as the pixel where the voltage pulse drops below TH0 last, as the time the voltage pulse is above TH0 is proportional to the energy deposited on the pixel.
%Counter 0 on the pixel which records the highest energy deposit, is also incremented, even if none of the summed voltage pulses at its summing nodes exceed TH1.
%The counts recorded by counter 0 are known as ``single pixel arbitrated'', and can be read out in addition to or instead of the counts recorded by counter 1.
%In the present work, we have characterised the performance of the GaAs:Cr and Si Medipix3 detectors operating in SPM and CSM, refraining from investigating the use of single pixel arbitrated as its effects are very similar to those obtained by operating in CSM and setting TH1 just above TH0.

The vast majority of room temperature semiconductor sensor materials are binary or ternary compounds, such as GaAs:Cr, CdTe and CdZnTe. 
The growth and fabrication of sensors from these materials is challenging and the presence of crystal defects and impurities can be common. 
These issues can lead to challenges such as incomplete charge collection, polarisation due to the build-up of trapped charge, high leakage currents, electric field instabilities and limitations in the sensor volumes that can be produced \cite{Tyazhev2003, Tlustos2008, Schlesinger2001}. 
In recent years, however, technologies for manufacturing high-Z sensors that can operate at room temperature and for bonding these to ASICs have matured. 
The GaAs:Cr material that we have characterised represents a significant step forward compared to earlier forms of semi-insulating GaAs that have been investigated for use in imaging detectors \cite{Tyazhev2013, Tyazhev2014}.
It has been used with a variety of ASICs, including the Medipix3 ASIC, for X-ray imaging \cite{Hamann2015a, Veale2014, Becker2017} and has been shown to be sufficiently robust for high-flux X-ray imaging \cite{Veale2014a}. 

In the case of the GaAs:Cr device that we have characterised, the sensor was bonded to the ASIC using a cold-weld indium bump-bonding process \cite{Schneider2015}, and had a Ni front-side contact.
Due to the electron mobility-lifetime product of GaAs:Cr being better than the hole mobility-lifetime product \cite{Veale2014}, the ASIC was set to collect electrons, and an applied negative bias of 300V was sufficient to to ensure complete collection of the electrons without any being lost due to trapping \cite{Hamann2015a}.
The Si detector consisted of high resistivity n-type Si with p+ on n implants that was bump-bonded using a standard solder bump-bonding process with an Al frontside contact. 
The Si sensor was operated with an applied positive bias of 110V, with the ASIC set to collect holes. 
Both devices were cooled passively, and their typical operational temperatures were approximately 28$^{\circ}$C.

\section{Characterisation of Detector Performance}
\label{S:3}
% Matt has also commented on this, take into account his suggestion too.
The performance of an imaging detector can be characterised through the measurement of independent and dependent parameters. 
Independent, directly measurable quantities include the MTF, which quantifies the ability of a detector to transfer contrast in an image as a function of spatial frequency, $\omega$.
This is the Fourier space representation of a detector's point spread function (PSF), which describes the detector's average response to an idealised input signal with the form of a $\delta$-function.
The noise power spectrum (NPS) is another independent parameter that measures how a detector transfers the noise present in the image incident upon it. 
Dependant on both of these quantities, the detector's DQE describes the extent to which that detector reproduces the signal-to-noise ratio (SNR) of features in the images it records as a function of their spatial frequency. 

For pixelated, digital detectors, it is necessary to distinguish between the presampling and digital forms of these measures of detector performance.
The presampling versions describe the detector's response without the effects of discrete sampling by pixels.
In HPDs, they describe how noise and signal are affected by the interactions of the primary electron with the sensor due to the scatter of the primary electron itself, the production of any secondary X-rays or electrons, and by the lateral spread of signal-carriers produced by the primary electron and any secondary radiation as they travel to the pixel electrodes.
%More generally, 
The presampling forms also account for integration over the effective pixel area.
In an ideal detector that counts all incident electrons only in the entry pixel, the effective pixel area is equivalent to the physical pixel area. 
However, it can be smaller than the physical pixel if electrons are not counted when they enter the sensor in certain regions of a pixel (e.g. the corner or edge) or greater than the physical pixel if multiple pixels count an incident electron.
The ratio of the effective pixel area to the physical pixel area is the detector's fill factor.

The digital MTF, NPS and DQE are their respective presampling form evaluated at the centre of each pixel \cite{Cunningham}. 
The finite size of the pixels means that aliasing of the digital MTF and NPS is possible, due to undersampling, causing them to be overestimated at high $\omega$ \cite{Metz1979}.
Various approaches have been used to determine the MTF and DQE of imaging detectors for use in electron microscopy  \cite{McMullan2007, Mcmullan2009, Meyer2000}, in part because it is necessary to treat different types of pixelated detectors in different ways. 
In the interests of ensuring our results can be readily compared with those characterising other imaging detector types, and with a view to clarifying what constitutes best practice in performing these measurements for HPDs in electron microscopy, we outline our approach for calculating the MTF, NPS and DQE and the justifications for it.
%The presampling MTF can be modelled as the convolution of the spatial distribution of incident electrons with the detector's aperture function, which describes the integration of electrons on each pixel.
%In the case of an idealised detector, incident electrons are counted only in the pixel in which they enter the sensor, and the aperture function is a sinc function

The presampling MTF can be measured directly if the experimental method used oversamples the input signal that approximates a $\delta$-function incident on the detector.
We have used the well-established knife-edge method \cite{Meyer2000}, informed by the approaches other authors have used when applying the technique to HPDs \cite{Tinti2018, McMullan2007}. 
In our measurements, we used a thick Al knife-edge positioned directly in front of the detector to minimise both geometric blurring and, as HPDs are sensitive to X-rays, the quantity of detectable X-rays produced by interactions of incident electrons with the edge. 
The knife-edge was set at an angle relative to the pixel columns such that the transition from the obscured portion of the pixel columns to the illuminated could be oversampled.
A region of the knife-edge without defects 16 pixels wide was identified, and for each column of pixels perpendicular to the edge in this region, the knife-edge location was identified with sub-pixel accuracy via interpolation as the position at which the intensity is half the column's maximum value.
Pixel values were then rearranged in order of their distance from the position of the knife-edge to give a single, oversampled edge-spread function (ESF).
The ESF can be differentiated directly to find the detector's line-spread function (LSF), which is equivalent to the PSF in one dimension, or it can be fitted with a function to minimise the effects of noise in the measurement.
We have found that a single error function, as defined in equation \ref{eqn:esf_fit},  provided a good fit to the ESF.
The MTF was then calculated as the modulus of the Fourier transform of the LSF that is calculated from differentiating $\textrm{ESF}_{fit}$.

\begin{equation}
    \centering
    \textrm{ESF}_{fit}(x)=\frac{\textrm{A}}{2}\bigg(1+\textrm{erf}~\bigg(\frac{\mu-x}{\sigma}\bigg)\bigg)  
    \label{eqn:esf_fit}
\end{equation}

In equation \ref{eqn:esf_fit}, $\mu$ is the mean position of the function, which is set to 0, $\sigma$ is the width of the error function and A is a normalisation factor. 
Some studies include a term to correct for the effects of integrating over the physical pixel area or apply a correction directly to the MTF \cite{Meyer2000}, but this has been shown to have a minimal effect on the final MTF \cite{Mcmullan2009} and was found to yield no significant improvement in the fit of our ESF. 

%DMcG - as written it's not clear that the following paragraph considers calculation of the digital form of the MTF - effectively we are saying that we did not take this step.
% KAP: I think a more critical issue is that I state earlier that the presampling MTF accounts for integration over the effective pixel size. The point I'm making about the effective pixel size is therefore nonsensical, and leads to confusion as to whether I'm discussing the presampling or digital MTF (I'm discussing the presampling MTF). 

%{\color{purple}We note that in the case of counting HPDs, where the effective pixel size is threshold-dependent, that the correction needed to account for the effects of integrating over the pixel would also be threshold dependent.
%This makes it impracticable to apply a correction to account for the effects of pixel integration, unless the effective pixel area is known independently.}
%Furthermore, as the effect of applying a counting threshold on detector performance and the threshold-dependence of this effect is of interest, it is not desirable to attempt to apply any such correction.

It is not possible to oversample the noise profile of the detector and identify the aliased contributions from above the detector's Nyquist frequency ($\omega_{N}$), which is its maximum sampling frequency and equal to $1/(2\times \textrm{pixel pitch})$. 
Consequently, the presampling NPS cannot be recovered.
%An explicit definition of the digital NPS, given in equation \ref{eqn:NPS_digital} therefore includes a term that reflects these contributions due to alaising. 
%DMcG - the following paragraph is a relatively long way of saying that we just can't effectively apply any correction to yield NPS_pre and so we will use NPS_dig - can it be reduced in length but still show that you have considered these points?
A correction for aliasing has been proposed for CCD cameras \cite{Meyer2000}.
Undersampling and aliasing are inevitable for scintillator-coupled CCD cameras as electrons can be registered by multiple pixels in spite of the CCD pixel fill factor being $<$~1 \cite{Zhao1997}, but this is not true for all pixelated detectors.  
The effective pixel area of HPDs can be larger than the physical pixel as incident electrons can be counted by multiple pixels, and they can therefore have an effective fill factor $>$ 1. 
% Note to self - you don't specify HPDs in your earlier discussion, so you would need to change wording here if you keep this sentence. 
This has the effect of an anti-aliasing filter \cite{Mcmullan2009}.
%and can have a beneficial effect on the DQE, as the high-frequency NPS can be suppressed to a greater extent than the high-frequency presampling MTF is, resulting in an enhanced DQE at high spatial-frequencies.
%It is therefore not desirable to apply an aliasing correction when using a low threshold
When using a high threshold, such that each incident electron is counted by at most one pixel, the NPS should be independent of spatial frequency and it is not necessary to account for aliasing. 
%Although it follows from this that a correction for aliasing would be valid at intermediate thresholds where electrons are counted by more than one pixel (or a pixel other than the one in which they enter the sensor) but any anti-aliasing effect is negligible, in practice identifying these thresholds and treating them separately is not practical.
%Additionally, in the case of the Medipix3's CSM, there is an additional noise component at high spatial frequencies due to the operation of the CSM algorithm over $2~\times~2$ pixel blocks that cannot be easily distinguished from the effects of aliasing. 
%In light of the number of cases where it is undesirable, unnecessary or unclear how to apply an correction for aliasing, we have not applied any such correction.
We have therefore not attempted to apply any correction for aliasing when calculating the NPS.
%\begin{equation}
%    \textrm{NPS}_{dig}(\omega) = \textrm{NPS}_{d}(\omega)+\Sigma^{\infty}_{m=1}\textrm{NPS}_{d}\bigg(\omega \pm \frac{m}{x_{0}}\bigg)
%    \label{eqn:NPS_dig}
%\end{equation}

\begin{equation}
    \centering
    \textrm{NPS}_{dig}(\omega_{x}, \omega_{y}) = \frac{x_{0}y_{0}}{\textrm{N}_{x}\textrm{N}_{y}}E\big\{|\textrm{FT}(\Delta d_{n_{x}, n_{y}})|^{2}\big\}
    \label{eqn:NPS}
\end{equation}

Equation \ref{eqn:NPS} provides a practical definition of the digital NPS of an imaging detector used in our calculations \cite{Cunningham}. 
N$_{x}$ N$_{y}$ are the number of pixels in the detector's x and y-axis respectively, while $x_{0}$ and $y_{0}$ are the pixel pitch in x and y respectively.
The value represented by $\Delta d_{n_{x}, n_{y}}$ is the difference between the number of counts registered by a pixel with coordinates ($x, y$) given a mean dose per pixel $n$ and the expectation value of the number of counts recorded by the pixel.
This was measured by recording a series of flat field exposures, calculating their mean image and subtracting this from each frame in the series.
The Fourier transform of each image was then calculated, and these were averaged to find the expectation value of the Fourier transform of $\Delta d_{n_{x}, n_{y}}$.
The 1D digital NPS was then found by taking the radial average. 

\begin{equation}
    \centering
    \textrm{DQE}_{dig}(\omega) = \frac{d_{n}^{2}\textrm{MTF}_{pre}^{2}(\omega)}{n~\textrm{NPS}_{dig}(\omega)}
    \label{eqn:DQE1}
\end{equation}

The digital DQE is defined by equation \ref{eqn:DQE1}, where $d_{n}$ is the mean number of counts recorded per pixel and $n$ is the mean number of electrons per pixel in the flat field images used to calculate the NPS. 
To find $n$, the beam current $I$ is measured and a series of images with a frame time $t$ are recorded, with the entirety of the beam incident on the detector.
The detector gain factor, $g$, is calculated using equation \ref{eqn:gain}, where $d_{n_{m}}$ is the mean number of counts recorded by the $m$-th pixel and $N$ the total number of pixels.
%can be found by measuring the beam current, $I$, and then spreading the beam so that it just covers the area of the sensor, estimating the area of the beam on the detector and hence the total does to the detector and then calculating the average dose to each pixel.
%In practice, it is easier to measure the detector's gain factor, $g$, the ratio of counts recorded to incident electrons by measuring the beam current and then recording images with the entire beam incident upon the detector with a frame time of $t$. 
%Equation \ref{eqn:gain}, where d$_{n_{m}}$ is the number of counts recorded by the $m$-th pixel and $N$ the total number of pixels, can then be used to calculate the value of $g$.
The value of $n$ for the flat field exposures used to determine the NPS was then found by calculating $d_{n}$ for these images using the fact that $n~=~d_{n}/g$.

\begin{equation}
    \centering
    g = \frac{\sum_{m=0}^{m=N-1} d_{n_{m}}~e}{I~t}
    \label{eqn:gain}
\end{equation}

A difficulty encountered in calculating the NPS in this way is that the low-frequency NPS tends to be incorrectly estimated \cite{Mcmullan2009}.
The standard solution to this is to normalise the NPS by NPS(0) and calculate DQE(0) separately, using this as a scaling factor.
This is formalised in equation \ref{eqn:DQE2}, where NNPS is the normalised NPS.
DQE(0) is calculated using equation \ref{eqn:DQE0}, where NPP is the noise-per-pixel, the corrected value of NPS(0). 
In principal, NPS(0) should be the variance, $\sigma_{d_{n}}^{2}$, of $\Delta d_{n_{x},n_{y}}$.
However, as electrons scatter over multiple pixels, there are correlations in the number of counts recorded by each pixel and the variance is not an accurate measurement of NPS(0).
%When the detector is operating in CSM, there are, as discussed previously, correlations in the noise across the pixels that work together to localise incident electrons. 
%The resulting high-frequency noise means the variance overestimates the value of NPS(0).
To find the NPP, the images of $\Delta d_{n_{x}, n_{y}}$ were binned by progressively larger factors, $b$ and the variance of the images, normalised by the square of the binning factor was evaluated. 
As $b$ increases, $\sigma_{d_{n}}^{2}/b^{2}$ reaches a plateau as the correlations between neighbouring pixels are discounted, which is taken to be the NPP.

\begin{equation}
    \centering
    \textrm{DQE}_{dig}(\omega) = \textrm{DQE}_{dig}(0)\frac{\textrm{MTF}_{pre}^{2}(\omega)}{\textrm{NNPS}_{dig}(\omega)}
    \label{eqn:DQE2}
\end{equation}
 
\begin{equation}
    \centering
    \textrm{DQE}(0) = \frac{d_{n}^{2}}{n~\textrm{NPP}}
    \label{eqn:DQE0}
\end{equation}

%As DEDs can easily distinguish between incident electrons and background noise, their performance can also be investigated by studying their response to individual electrons. 
%In the case of counting HPDs, the use of a threshold to discount detector noise makes this particularly easy. 
%Frames recorded with short acquisition times with a low electron dose show only pixels of clusters due to individual electrons. 
%These can be analysed to understand the probabilities of different types of electron interactions, as indicated by different pixel cluster shapes, as well as the detector's PSF. 
%To investigate the effects of varying the counting threshold on the response of the Si and GAs:Cr detectors to individual electrons, we have varied the value of TH0 and TH1 when the detectors were operating in SPM and CSM respectively.
%At each threshold, 2000 frames with a 50ns acquisition time were recorded, to ensure good statistics for the purpose of analysis although a low dose rate was used.
%When performing these measurements, 

Measurements using electrons with energies in the range of 60~-~200keV for both Si and GaAs:Cr detectors were performed by mounting the detectors on an FEI Tecnai T-20 TEM using the 35mm port above the viewing screen. 
The beam current for these measurements was recorded using a Faraday cup mounted at the end of the chassis containing the detector, which was connected to a Keithly 485 Picoammeter.
To acquire 300keV electron data, the GaAs:Cr detector was mounted in the Gatan camera block on a FEI Titan 80~-~300 (S)TEM.
For this set of measurements, the beam current was calculated using the number of counts registered by a Gatan Ultrascan located in the same plane as the Medipix3 device and the manufacturer-provided conversion factor.
When acquiring flat field, knife-edge and gain images, 128 frames at each threshold were recorded, with the frame time varied to maximise the number of counts for a particular beam current without saturating the on-pixel counters.
It should be noted that the DQE can be sensitive to electron flux and dose. 
If the flux is too high then the detector is unable to count all incident electrons, which causes a suppression of the NPS at low spatial frequencies \cite{Li2013}, but if the value of $n$ is too low then the NPS will be dominated by shot noise in the incident number of electrons \cite{Kuijper2015}.
In our measurements we have taken care to use beam currents that were sufficiently low so as to minimise undercounting of incident electrons while also ensuring that the value of $n$ is statistically significant.

The detector thresholds, TH0 and TH1 when operating in SPM and CSM respectively, were calibrated using fluorescence X-rays from a series of targets. 
This provides an absolute energy calibration, as low-energy photons typically deposit their energy in a single interaction, rather than scattering over multiple pixels as is the case for high-energy electrons.
Consequently, the disparity between the maximum amount of energy deposited in a pixel by incident electrons and the initial energy of incident electrons is apparent.
As the energy of the incident electrons increases, the maximum amount of energy deposited on a single pixel as a fraction of the primary electron energy decreases, due to increased scatter. 
This effect is more pronounced when in SPM, as in CSM the energy deposited over 2~$\times$~2 pixel blocks by incident electrons is summed.
A result of this is that the maximum threshold at which it was possible to fit the knife-edge data with equation \ref{eqn:esf_fit} is lower than the counting threshold that corresponds to the primary electron energy, substantially so for electrons with energies $\geq$~120keV. 
For example, for the 300keV electron data acquired with a GasAs:Cr device, the maximum amount of energy deposited on a single pixel when the detector was working in SPM was 160keV, and the highest threshold at which it was possible to fit equation \ref{eqn:esf_fit} was lower than this (131.3keV), due to insufficient counts being recorded at thresholds close to 160keV.

\section{MTF and DQE Measurements}
\label{S:4}

\begin{figure}[h!]
    \centering
    \includegraphics[scale=0.3]{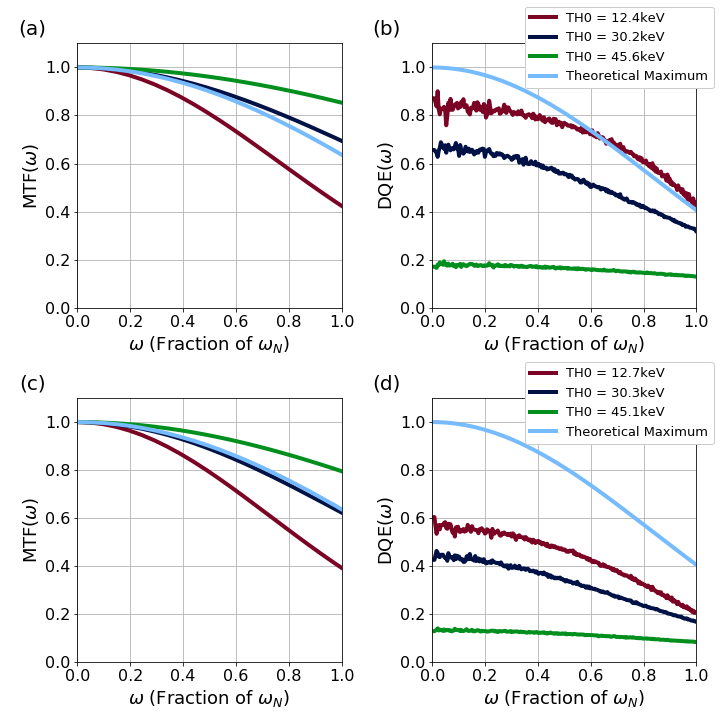}
    \caption{(a) MTF and (b) DQE measurements using selected thresholds for a Si device operating in SPM for 60keV electrons; (c) MTF and (d) DQE measurements for a GaAs:Cr detector under the same conditions.}
    \label{fig:60keV_SPM_MTF_DQE}
\end{figure}

\begin{figure}[h!]
    \centering
    \includegraphics[scale=0.3]{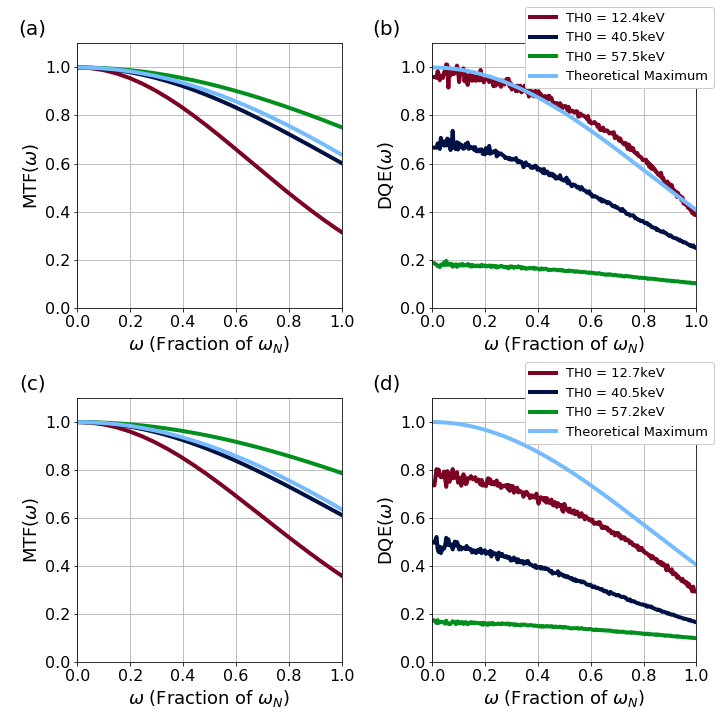}
    \caption{MTF for (a) Si and (c) GaAs:Cr detectors operating in SPM with selected counting thresholds for 80keV electrons; the corresponding DQE results for (b) a Si device and (d) a GaAs:Cr device.}
    \label{fig:80keV_MTF_DQE}
\end{figure}

Figures \ref{fig:60keV_SPM_MTF_DQE} and \ref{fig:80keV_MTF_DQE} show MTFs and DQEs obtained using the lowest threshold above both detectors' noise levels; a threshold equal to half the primary electron energy and the highest threshold common to both devices at which the knife-edge data could be fit with equation \ref{eqn:esf_fit} for 60keV and 80keV electrons respectively. 
%At these low electron energies, there is minimal difference between the two devices in terms of MTF, but significant difference in DQE. 
The MTFs of the Si device in figure \ref{fig:60keV_SPM_MTF_DQE}(a) are slightly superior to those of the GaAs:Cr device in \ref{fig:60keV_SPM_MTF_DQE}(c) for a given threshold, with the greatest difference being 0.07 at $\omega_{N}$ for the intermediate threshold MTF.
In figure \ref{fig:60keV_SPM_MTF_DQE}(b), the Si DQEs are significantly higher than their GaAs:Cr counterparts in 1(d), with the largest difference being 0.34 between the low threshold DQEs at low $\omega$.
Comparing the MTFs of the Si detector for 80keV electrons in figure \ref{fig:80keV_MTF_DQE}(a) with those of the GaAs:Cr detector in \ref{fig:80keV_MTF_DQE}(c), the performance of the latter is marginally better, with the greatest difference being that the low threshold GaAs:Cr MTF is 0.05 higher than the Si MTF at $\omega_{N}$.
However, the DQEs of the Si detector seen in figure 2(b) are again significantly higher than those of the GaAs:Cr detector in 2(d), with the greatest difference between the GaAs:Cr and Si DQEs at low and intermediate thresholds being at least 0.26.

An ideal detector counts all incident electrons only in the entry pixel, and such a detector would have a constant NPS, unitary gain and, if it had square pixels, a PSF with the form of a top-hat function, a constant NPS and unitary gain.
These factors give rise to an ideal MTF that is a sinc function equal to 0.64 at $\omega_{N}$ and, per equation \ref{eqn:DQE1}, an ideal DQE that is the square of the ideal MTF, with a value of 0.41 at $\omega_{N}$ \cite{Ruskin2013}.
These are plotted in figures \ref{fig:60keV_SPM_MTF_DQE} and \ref{fig:80keV_MTF_DQE} for comparison with the experimental results.
A notable feature of the DQE results for the Si detector in figures \ref{fig:60keV_SPM_MTF_DQE}(b) and \ref{fig:80keV_MTF_DQE}(b) is that, when using a low threshold, the detector is able to surpass the ideal DQE of a pixelated detector.
At 60keV, its best DQE is at least 0.1 less than the theoretical maximum when $\omega \leq 0.1\omega_{N}$ , but it surpasses the theoretical maximum at high $\omega$, while at 80keV the low threshold DQE of Si detector surpasses the ideal at intermediate values of $\omega$.
This can be attributed to the fact that, while both the MTF and NPS are suppressed at high spatial frequencies when an electron is counted by multiple pixels, the NPS is suppressed to a greater extent relative to the MTF, inflating the DQE at high spatial frequencies so that it exceeds the predicted DQE of an ideal detector \cite{Mcmullan2009}.

\begin{figure}[h!]
    \centering
    \includegraphics[scale=0.3]{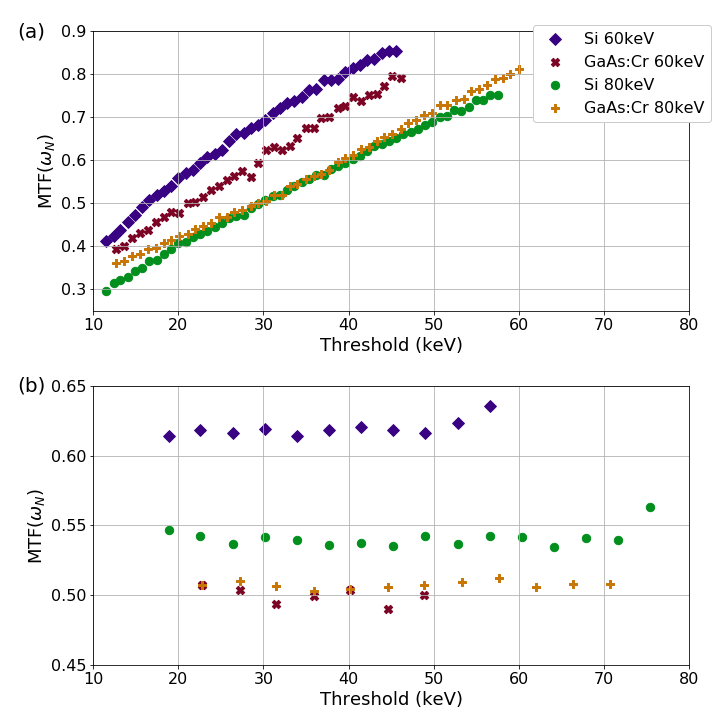}
    \caption{MTF($\omega_{N}$) for Si and GaAs:Cr devices as a function of threshold for 60keV and 80keV electrons as a function of counting threshold when operating in (a) SPM and (b) CSM.}
    \label{fig:LowEnergy_MTF_Nyquist}
\end{figure}

In figure \ref{fig:LowEnergy_MTF_Nyquist} MTF($\omega_{N}$) as a function of threshold for both detectors when operating in SPM and CSM for 60keV and 80keV electrons is plotted. 
For both detectors in SPM, the value of MTF($\omega_{N}$) increases approximately linearly with the counting threshold. 
This is consistent with the expectation that the effective pixel area decreases with increasing counting threshold \cite{Mir2017, McMullan2007}. 
At low thresholds, the effective pixel size is larger than the physical pixel pitch and a pixel can count an incident electron even if the electron is not incident on that pixel, causing a reduction in the MTF.
The higher the counting threshold, the more energy an electron must deposit in a pixel to be counted by that pixel, decreasing the effective pixel size. 
Consequently, the value of MTF($\omega_{N}$) can exceed the theoretical maximum of 0.64 for an ideal detector at high thresholds, as the effective pixel size can be smaller than the physical pixel pitch of the detector.

The 60keV results for the two devices operating in SPM in figure \ref{fig:LowEnergy_MTF_Nyquist}(a) confirm that the the Si device consistently outperforms the GaAs:Cr device across all counting thresholds, with there being an almost constant offset between the two curves. 
However, the difference is small, with a maximum value of 0.08 when the GaAs:Cr and Si detector thresholds are both 28.5keV.
At 80keV, the GaAs:Cr marginally outperforms the Si device at low and high counting thresholds, with the MTF($\omega_{N}$) values of the two devices overlapping at intermediate thresholds in figure \ref{fig:LowEnergy_MTF_Nyquist}(a). 
The maximum difference in MTF($\omega_{N}$) for 80keV electrons is also 0.08, when the GaAs:Cr and Si device thresholds are 58.1keV and 57.5keV.

When operating in CSM, the performance of two devices is very different, with MTF($\omega_{N}$) being independent of counting threshold in figure \ref{fig:LowEnergy_MTF_Nyquist}(b).
This follows from the fact that, when the devices are operating in CSM, whether or not an electron is counted depends on the sum of the charge induced in neighbouring pixels.
For low-energy electrons, which typically deposit all their energy across one of the 2~$\times$~2 pixel blocks that the CSM algorithm operates across, this will be consistently above threshold until the threshold is equal to the energy of the incident electron. 
The extent to which MTF($\omega_{N}$ deviates from the ideal value of 0.64 is indicative of how successful the CSM algorithm is at identifying the pixel of entry.
When operating in CSM, MTF($\omega_{N}$) for the Si detector with 60keV electrons is similar to that of an ideal detector, ranging between 0.61 and 0.64.
The response of the Si detector is poorer for 80keV electrons, ranging between 0.53 and 0.56, but this is still better than the performance of the GaAs:Cr device used in CSM, for which MTF($\omega_{N}$) fluctuates around 0.50 for 60keV electrons and around 0.55 for 80keV electrons. 

\begin{figure}[h!]
    \centering
    \includegraphics[scale=0.3]{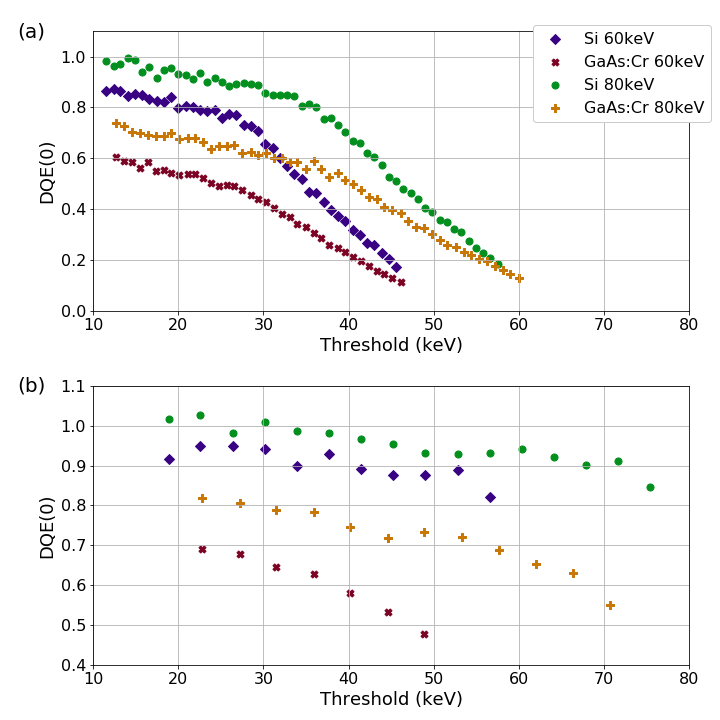}
    \caption{DQE(0) as a function of threshold for Si and GaAs:Cr detectors for 60keV and 80keV electrons in (a) SPM, (b) CSM.}
    \label{fig:LowEnergy_DQE_Zero}
\end{figure}

\begin{figure}[h!]
    \centering
    \includegraphics[scale=0.3]{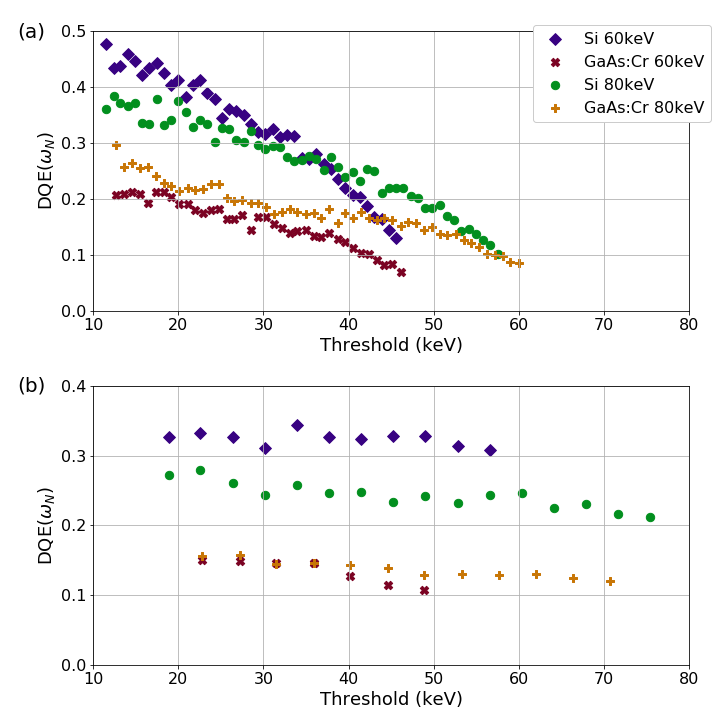}
    \caption{Dependence of DQE($\omega_{N}$) on counting threshold for GaAs:Cr and Si devices operating in (a) SPM and (b) CSM, for electrons with energies of 60keV and 80keV.}
    \label{fig:LowEnergy_DQE_Nyquist}
\end{figure}

The DQE(0) and DQE($\omega_{N}$) results for the two detectors in both modes of operation for 60keV and 80keV electrons are shown in figures \ref{fig:LowEnergy_DQE_Zero} and  \ref{fig:LowEnergy_DQE_Nyquist}.
These results make apparent the difference in performance between the two detectors when using electrons at these energies.
When the detectors operate in SPM, DQE(0) decreases gradually with increasing threshold up to approximately half the primary electron energy, but at thresholds above this, DQE(0) decreases rapidly as the counting threshold is increased.
DQE($\omega_{N}$) follows a similar trend, although the initial decrease with regard to threshold is greater for the Si DQE($\omega_{N}$).
For both detectors operating in CSM, DQE($\omega_{N}$) is approximately constant for both 60keV and 80keV electrons, with minimal negative dependence on threshold seen in figure \ref{fig:LowEnergy_DQE_Nyquist}(b).
The CSM DQE(0) in figure \ref{fig:LowEnergy_DQE_Zero}(b) exhibits a greater dependence on threshold, decreasing as the threshold increases for both devices and at both electron energies.
In the case of the GaAs:Cr device, it is possible to discern a similar trend to that seen in the SPM results in figure \ref{fig:LowEnergy_DQE_Nyquist}(a).

The negative dependence of the DQE on counting threshold seen in the SPM results arises from a similar reason as the positive dependence of MTF($\omega_{N}$) on counting threshold.
As the effective pixel size decreases with increasing threshold, so too does the likelihood of electrons not being registered by the detector, reducing the efficiency of the detector.
In theory, the value of DQE(0) should be constant for all counting thresholds up to half the primary electron energy, as it is only when the threshold is equal to half the primary electron energy that the effective pixel size should be less than the physical pixel size, leading to electrons not being counted.
In practice, as the electrons scatter through the sensor depositing their energy in multiple pixels, some electrons will not be registered at lower thresholds as they fail to deposit enough energy in any single pixel to be counted.

The relative constancy of DQE(0) and DQE($\omega_{N}$) as a function of threshold for the CSM results can be attributed to the same factors as the constancy of MTF($\omega_{N}$), namely that the signal recorded by neighbouring pixels is summed together.
Nevertheless, there is still a gradual decrease in the value of DQE(0) with increasing counting threshold.
This is most probably due to a small percentage of electrons that deposit their energy over an extended number of pixels greater than the 2~$\times$~2 blocks of pixels that the CSM algorithm works across being discounted, as no single block of pixels has the full energy of the electron deposited in it.

\begin{figure}[h!]
    \centering
    \includegraphics[scale=0.3]{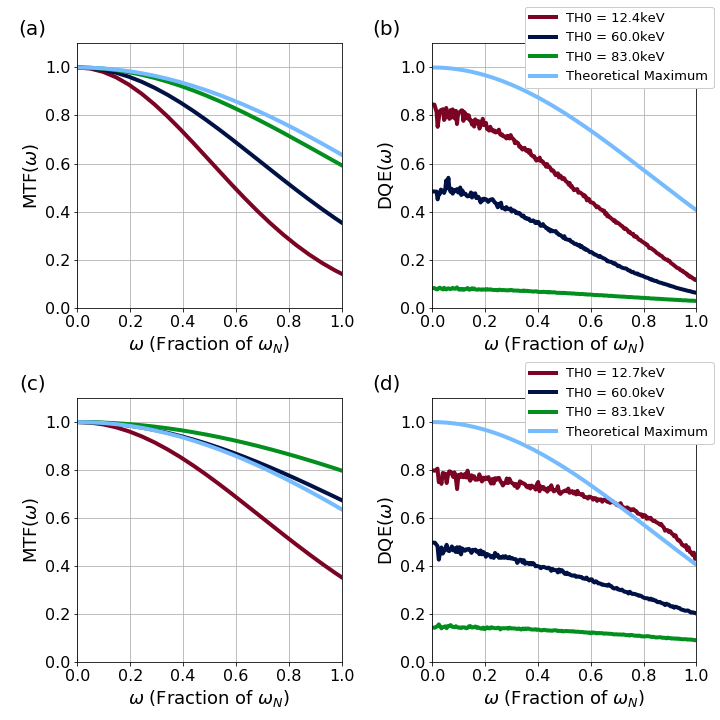}
    \caption{(a) MTF and (b) DQE at selected thresholds for a Si device operating in SPM for 120keV electrons; (c) MTF and (d) DQE for a GaAs:Cr detector in SPM for 120keV electrons at selected thresholds.}
    \label{fig:120keV_SPM_MTF_DQE}
\end{figure}

At increasing electron energy, the benefits of the GaAs:Cr sensor become apparent.
Figure \ref{fig:120keV_SPM_MTF_DQE} shows the MTFs and DQEs for the Si and GaAs:Cr devices in SPM at 120keV using the lowest and highest thresholds common to both detectors for which the knife-edge could be fit and the threshold closest to half the primary electron energy.
Both the MTFs and the DQEs of the GaAs:Cr device are superior to those of the Si detector.
At each threshold shown in figures \ref{fig:120keV_SPM_MTF_DQE}(a) and (c) the MTF of the GaAs:Cr device is at least 0.21 higher than the MTF of the Si device at $\omega_{N}$.
The DQE of the GaAs:Cr device at $\omega_{N}$ is 0.32 greater than the Si device for the lowest threshold in figures \ref{fig:120keV_SPM_MTF_DQE}(b) and (d), though this difference decreases to 0.14 and 0.06 for the intermediate and high thresholds.

\begin{figure}[h!]
    \centering
    \includegraphics[scale=0.3]{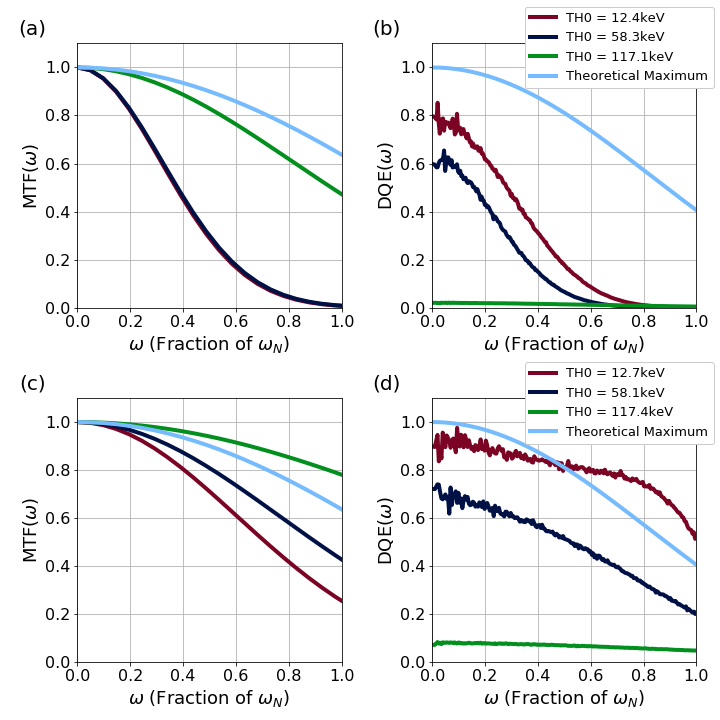}
    \caption{200keV SPM (a) MTF and (b) DQE at selected thresholds for a Si detector at selected thresholds and selected (c) MTF and (d) DQE for a GaAs:Cr device under the same conditions.}
    \label{fig:200keV_SPM_MTF_DQE}
\end{figure}

At 200keV the GaAs:Cr clearly outperforms the Si detector in both MTF and DQE.
Figure \ref{fig:200keV_SPM_MTF_DQE} shows the MTFs and DQEs of the two detectors operating in SPM for 200keV electrons using the lowest threshold above the noise level of both detectors; the highest  threshold  common  to both devices at which the knife-edge data could be fit and the threshold equal to half the highest threshold used for both devices.
The different choice in thresholds shown in figure \ref{fig:200keV_SPM_MTF_DQE} compared to figures \ref{fig:60keV_SPM_MTF_DQE}, \ref{fig:80keV_MTF_DQE} and \ref{fig:120keV_SPM_MTF_DQE} is due to the increased disparity between the maximum amount of energy deposited on a pixel and the primary electron energy when using higher-energy electrons discussed in section 3.
The MTF of the GaAs:Cr device is at least 0.25 greater than that of the Si detector at $\omega_{N}$ for all thresholds shown in figures \ref{fig:200keV_SPM_MTF_DQE}(a) and (c). 
At the low and intermediate thresholds the GaAs:Cr DQE is 0.52 and 0.21 higher than the Si DQE at $\omega_{N}$, while at the high threshold the difference in DQE is marginal.
Overall, the difference in performance between the GaAs:Cr and Si detectors is greater at 200keV than it is at lower electron energies for which the performance of the Si detector is comparable to or greater than that of the GaAs:Cr detector.

\begin{figure}[h!]
    \centering
    \includegraphics[scale=0.3]{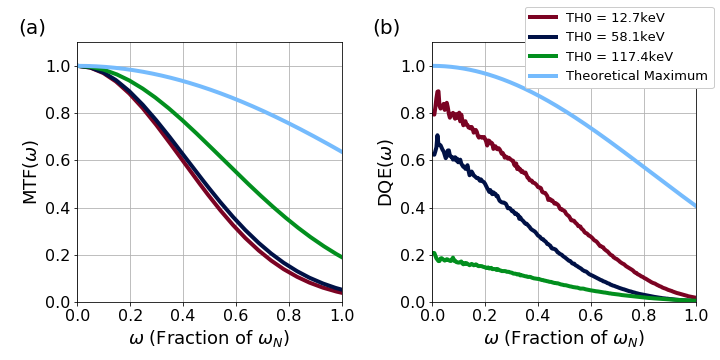}
    \caption{(a) MTF and (b) DQE at selected thresholds for a GaAs:Cr Medipix3 device operating in SPM for 300keV electrons.}
    \label{fig:300keV_GaAS_SPM_MTF_DQE}
\end{figure}

In figures \ref{fig:120keV_SPM_MTF_DQE}(d) and \ref{fig:200keV_SPM_MTF_DQE}(d), the low threshold GaAs:Cr DQE surpasses the DQE of an ideal detector at high $\omega$.
This is similar to the behaviour of the Si detector operating in SPM with 60keV and 80keV electrons in figures \ref{fig:60keV_SPM_MTF_DQE}(b) and \ref{fig:80keV_MTF_DQE}(b) and is caused by the same factors. 
While the extent to which the DQE of the Si detector for 60 and 80keV electrons surpasses the theoretical maximum is marginal, the DQE of the GaAs:Cr detector for 200keV electrons at high $\omega$ is significantly greater than that predicted for an ideal detector.
The greatest extent to which the low threshold DQE of the GaAs:Cr detector exceeds the DQE of an ideal detector is by 0.19 at a spatial frequency of 0.86$\omega_{N}$. 
At $\omega_{N}$, the low threshold DQE is equal to 0.52, which agrees closely with the maximum improvement in DQE that can be obtained due to the NPS being suppressed to a greater extent than the MTF \cite{Mcmullan2009}.

However, the performance of the GaAs:Cr detector deteriorates for 300keV electrons.
The MTFs and DQEs of the GaAs:Cr device for 300keV electrons using the same thresholds used in figure \ref{fig:200keV_SPM_MTF_DQE} are shown figure \ref{fig:300keV_GaAS_SPM_MTF_DQE}. 
These make apparent the degradation in performance of the GaAs:Cr device, being comparable to the the MTFs and DQEs of the Si device at 200keV in figures \ref{fig:200keV_SPM_MTF_DQE}(a) and (b). 

\begin{figure}
    \centering
    \includegraphics[scale=0.3]{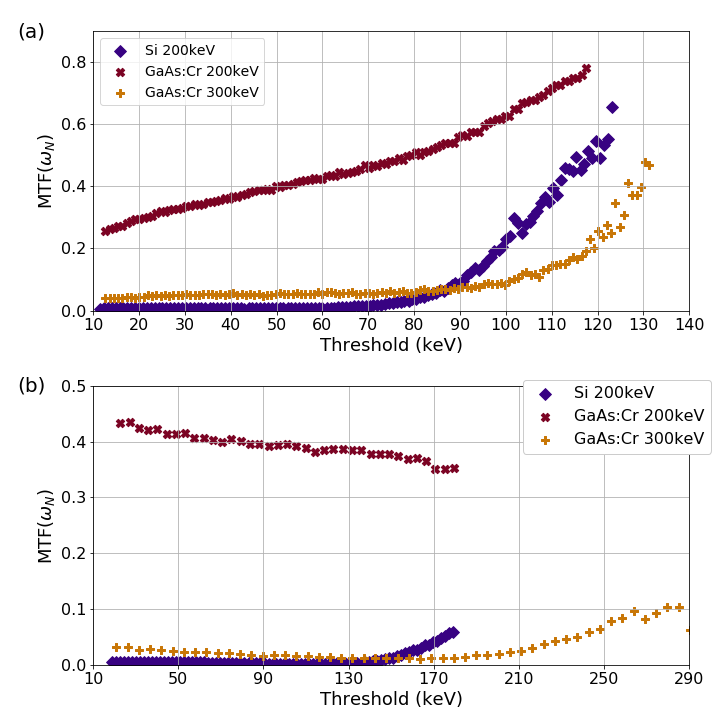}
    \caption{Dependence of MTF($\omega_{N}$) on counting threshold for a Si detector for 200keV electrons and a GaAs:Cr detector for 200keV and 300keV electrons in (a) SPM and (b) CSM.}
    \label{fig:HighEnergy_MTF_Nyquist}
\end{figure}

Examination of the 200keV and 300keV MTF($\omega_{N}$) as a function of threshold in figure \ref{fig:HighEnergy_MTF_Nyquist} confirms the similarity in performance between the Si detector for 200keV electrons and the GaAs:Cr detector for 300keV electrons. 
It also confirms the superiority of the GaAs:Cr detector MTF for 200keV electrons compared with the Si detector for both modes of operation. 
In figure \ref{fig:HighEnergy_MTF_Nyquist}(a), the value of MTF($\omega_{N}$) of the GaAs:Cr detector for 300keV electrons is 0.04 at low thresholds.
This is slightly higher than the Si detector's MTF($\omega_{N}$) of 0.01 at low threshold for 200keV electrons.
With increasing threshold, the response of the Si detector at 200keV surpasses the GaAs:Cr detector's response to 300keV electrons when they operate in SPM.
The maximum value of MTF($\omega_{N}$) for the Si detector is 0.66 at a threshold of 123.1keV, while that of the GaAs:Cr detector for 300keV electrons is 0.48 at a threshold of 130.4keV.
Although the Si detector in SPM is able to just outperform an ideal detector when using a high threshold, MTF($\omega_{N}$) for the GaAs:Cr device for 200keV electrons is consistently higher, with a minimum of 0.26 at a threshold of 12.7keV and a maximum of 0.78 at a threshold of 117.4keV.

Similarly, when the devices operate in CSM, MTF($\omega_{N}$) for the GaAs:Cr at 200keV is consistently better than MTF($\omega_{N}$) for the GaAs:Cr detector at 300keV and for the Si detector at 200keV.
In figure \ref{fig:HighEnergy_MTF_Nyquist}(b), the low threshold value of MTF($\omega_{N}$) for the Si detector for 200keV electrons is 0.01, while for the GaAs:Cr detector for 300keV electrons  MTF($\omega_{N}$) is 0.03 at the lowest threshold used.
The low threshold MTF($\omega_{N}$) for the GaAs:Cr detector for 200keV electrons is 0.43, and a difference between the detector's response in SPM and CSM is that MTF($\omega_{N}$) decreases with increasing threshold, reaching a minimum value of 0.35 at a threshold of 175.1keV.
For 200keV electrons with the Si detector operating in CSM, MTF($\omega_{N}$) decreases to a minimum of 0.00 at a threshold of 109.3kV before increasing at thresholds above 130keV to a maximum of 0.06 at a threshold of 179.1keV.
Likewise, MTF($\omega_{N}$) for the GaAs:Cr detector for 300keV electrons decreases to a minimum of 0.01 at a threshold of 163.5keV before increasing to a maximum of 0.1 at a threshold of 279.7keV.

The various different trends in MTF($\omega_{N}$) as a function of threshold for the two devices for 200keV and 300keV electrons in figure \ref{fig:HighEnergy_MTF_Nyquist} can all be attributed to the same principles that describe the low-energy electron MTF($\omega_{N}$), DQE(0) and DQE($\omega_{N}$) in figures \ref{fig:LowEnergy_MTF_Nyquist} - \ref{fig:LowEnergy_DQE_Nyquist}.
The range of 200keV electrons is sufficiently reduced in the GaAs:Cr detector so that the SPM MTF($\omega_{N}$) increases with increasing threshold, for the same factors that explain the low-energy SPM MTF($\omega_{N}$) dependence on threshold in figure \ref{fig:LowEnergy_MTF_Nyquist}(a).
The dependence of MTF($\omega_{N}$) on threshold for 200keV electrons and 300keV electrons for the Si and GaAs:Cr devices operating in SPM can be explained by a combination of the increased ranges of 200keV and 300keV electrons in Si and GaAs:Cr respectively and the tendency of electrons to deposit more energy towards the end of their trajectory in the sensor rather than at the beginning \cite{Ryll2019, Correa2020}.
Increasing the counting threshold does not initially improve the MTF (which is also apparent in figures \ref{fig:200keV_SPM_MTF_DQE}(a) and \ref{fig:300keV_GaAS_SPM_MTF_DQE}(a)), as using a high threshold is more likely to count the electron in a pixel at the end of the electron's trajectory rather than the entry pixel.
However, at very high thresholds, work with the Eiger detector \cite{Tinti2018} suggests that the only electrons counted are the small fraction of electrons that deposit most of their energy close to their entry point, causing an improvement in MTF.

Although 200keV electrons have a sufficiently reduced range in GaAs:Cr such that in SPM using a high threshold successfully identifies the entry pixel of those electrons that are counted, they are still able to deposit their energy over multiple 2~$\times$~2 pixel blocks.
Consequently, when the GaAs:Cr sensor operates in CSM, multiple pixels register 200keV electrons when using a low threshold.
As the counting threshold increases, hits associated with the pixel block that has the most energy deposited on it continue to be counted, but this block does not necessarily contain the entry pixel. 
Increasing the counting threshold therefore suppresses hits associated with the block containing the entry pixel, which causes the decrease in MTF($\omega_{N}$) with increasing threshold seen in figure \ref{fig:HighEnergy_MTF_Nyquist}.
This also explains the initial decrease seen in the 300keV GaAs:Cr CSM MTF($\omega_{N}$) and 200keV Si CSM MTF($\omega_{N}$).
The increase in MTF($\omega_{N}$) at high threshold for both 300keV electrons and the GaAs:Cr detector and 200keV electrons and the Si detector in CSM occurs for the same factors as the improvement in their SPM counterparts at high threshold.

\begin{figure}[h!]
    \centering
    \includegraphics[scale=0.3]{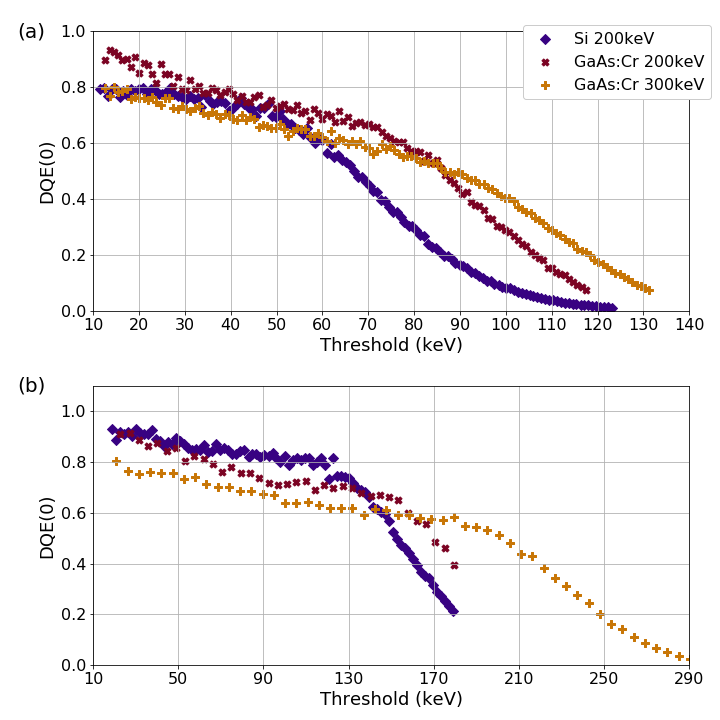}
    \caption{DQE(0) as a function of counting threshold for a Si device with 200keV electrons and a GaAs:Cr device for 200keV and 300keV electrons with the devices operating in (a) SPM (b) CSM.}
    \label{fig:HighEnergy_DQE_Zero}
\end{figure}

\begin{figure}[h!]
    \centering
    \includegraphics[scale=0.3]{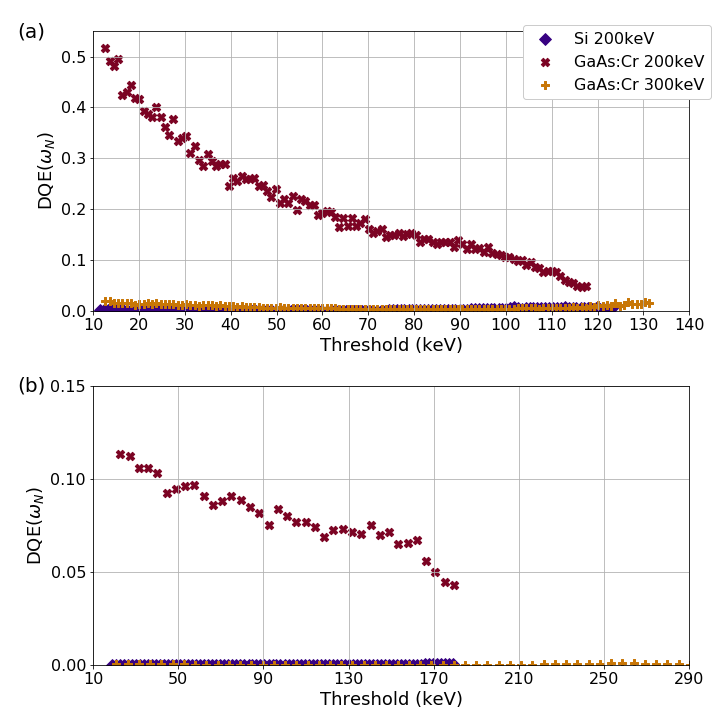}
    \caption{DQE($\omega_{N}$) for a GaAs:Cr device for 200keV and 300keV electrons and a Si device for 200keV electrons in (a) SPM and (b) CSM as a function of threshold.}
    \label{fig:HighEnergy_DQE_Nyquist}
\end{figure}

In figure \ref{fig:HighEnergy_DQE_Zero}, the trends in DQE(0) as a function of threshold for both devices are consistent with the interpretation of the MTF($\omega_{N})$ data above.
Operating in SPM, the 200keV DQE(0) for the Si and GaAs:Cr detectors and the 300keV DQE(0) for the GaAs:Cr detector in figure \ref{fig:HighEnergy_DQE_Zero}(a) exhibit a similar trend as to that seen for the low-energy SPM results in figure \ref{fig:LowEnergy_DQE_Zero}(a). 
The principle difference between the results in figures \ref{fig:LowEnergy_DQE_Zero}(a) and \ref{fig:HighEnergy_DQE_Zero}(a) is that in the latter the threshold at which there is a change in gradient in the dependence of DQE(0) on threshold is approximately half the maximum energy deposited by the primary electron on a single pixel, rather than approximately half the primary electron energy as in figure \ref{fig:LowEnergy_DQE_Zero}(a).
For 200keV electrons, the Si detector DQE(0) is 0.80 at a threshold of 12.4keV, while the GaAs:Cr detector DQE(0) for 200keV electrons is 0.90 at a threshold of 12.7keV. 
At the same threshold but for 300keV electrons, DQE(0) of the GaAs:Cr detector is 0.80.
The Si detector 200keV DQE(0) falls to 0.01 at a threshold of 123.1keV, while the GaAs:Cr detector DQE(0) has a minimum of 0.07 at a threshold of 117.4keV for 200keV electrons and a minimum of 0.08 at a threshold of 131.3keV for 300keV electrons.
The difference seen in the rate at which DQE(0) decreases with increasing threshold for the two detectors for 200keV electrons can be attributed to the spread of the signal produced by 200keV electrons being greater in the Si detector than it is in the GaAs:Cr detector.
Consequently, more electrons deposit enough energy on a single pixel in the GaAs:Cr sensor to be counted for a given threshold than they do in the Si detector.
That the high threshold value of DQE(0) for 300keV electrons in the GaAs:Cr sensor is greater than the high threshold DQE(0) for 200keV electrons in either device is due to both the reduced spread in signal in the GaAs:Cr sensor and the fact that the net energy that can be deposited by a 300keV electron is greater than that deposited by a 200keV electron.

%That the highest viable threshold used for both detectors at high electron energies and when operating in SPM is proportional to the primary electron energy, is much lower that the highest usable threshold for the low-energy measurements also indicative of the increased lateral spread in the signal generated by primary electrons of higher energies compared to their low-energy counterparts in both sensors. 

When the detectors operate in CSM, there is a notable difference between the dependence of DQE(0) on threshold at high electron energies compared with at low electron energies.
For the CSM DQE(0) in figure \ref{fig:HighEnergy_DQE_Zero}(b), there is a marked increase in the rate at which the value of DQE(0) decreases at high thresholds relative to the rate at which it decreases at low threshold.
This is similar to the dependence of DQE(0) on threshold when the detectors operate in SPM, rather than the gradual decrease with increasing threshold seen at 60 and 80keV (figure \ref{fig:LowEnergy_DQE_Zero}(b)), where any change in gradient is slight and difficult to discern.
The difference between the low-energy and high-energy trends can be explained by the same factors that explain the dependence of MTF($\omega_{N}$) on threshold in figure \ref{fig:HighEnergy_MTF_Nyquist}(b).
At low thresholds, the number of counts decreases gradually with respect to threshold, as the incident electrons deposit their energy over multiple CSM pixel blocks and, in each block, the reconstructed charge needs to be above threshold for it to be counted.
Above a certain counting threshold, electrons are only counted in a single pixel block and as the counting threshold increases the greater the energy deposited in that block must be for the electron to be counted.
This is analogous to the detector behaviour in SPM, but with the added complication of how the electron energy is deposited over blocks of neighbouring pixels rather than in single pixels. 
The thresholds at which the gradient of DQE(0) as a function of counting threshold changes for 200keV electrons and the Si detector and 300keV electrons and the GaAs:Cr detector are also the thresholds above which the corresponding MTF($\omega_{N}$) begin to rapidly increase with respect to threshold, corroborating the interpretation of the high threshold improvement in MTF in figure \ref{fig:HighEnergy_MTF_Nyquist}(b).

DQE($\omega_{N}$) as a function of counting threshold for the two detectors operating in SPM and CSM for high-energy electrons is shown in figure \ref{fig:HighEnergy_DQE_Nyquist}.
Consistent with the results in figures \ref{fig:HighEnergy_MTF_Nyquist} and \ref{fig:HighEnergy_DQE_Zero}, the 200keV DQE($\omega_{N}$) of the GaAs:Cr detector is  significantly better the 300keV DQE($\omega_{N}$) of the GaAs:Cr detector and the Si detector 200keV DQE($\omega_{N}$) in both modes of operation.
In SPM, the maximum GaAs:Cr DQE($\omega_{N}$) for 200keV electrons is 0.52 at a threshold of 12.7keV.
This decreases with threshold to a minimum value of 0.05 at a threshold of 116.5keV.
The value of DQE($\omega_{N}$) at 200keV for Si is 0.00 at low threshold, increasing to 0.01 at thresholds greater than 106.9keV.
The trend seen in the 300keV GaAs:Cr SPM DQE($\omega_{N}$) is noteworthy. 
At a threshold of 12.7keV it is equal to 0.02 and with increasing threshold decreases to a minimum of 0.00 at a threshold of 79.4keV before increasing to 0.02 again at 131.3keV.
This behaviour reflects the fact that increasing the counting threshold initially does little to improve the MTF but does increase undercounting of incident electrons, but at very high thresholds the improvement in MTF at high $\omega$ outweights the adverse impact of undercounting electrons on the DQE.
The CSM DQE($\omega_{N}$) in figure \ref{fig:HighEnergy_DQE_Nyquist}(b) confirms the failure of the CSM algorithm to enhance detector performance for high-energy electrons in both sensors. 
At low thresholds, the CSM DQE($\omega_{N}$) of the GaAs:Cr detector is lower than the low threshold SPM DQE($\omega_{N}$), with a maximum value of 0.11 at a threshold of 22.8keV.
This decreases to 0.04 at a threshold of 179.4keV.
The Si detector CSM DQE($\omega_{N}$) for 200keV electrons never exceeds 0.00 for all thresholds, and this is also true of DQE($\omega_{N}$) for the GaAs:Cr detector in CSM for 300keV electrons.
%Close examination of the MTF($\omega_{N}$) shown in figure \ref{fig:HighEnergy_MTF_Nyquist}b confirms that the same trend is seen for the 300keV GaAs:Cr CSM MTF($\omega_{N}$) data.
%This is likely due to high-energy electrons tending to deposit the majority of their energy in a pixel that is not the entry pixel. 
%Consequently, there is initially a decrease in the Nyquist value of the MTF and DQE as the entry pixel tends to discounted because the likelihood of it counting the incident electron decreases with increasing threshold.
%At very high thresholds, the only events registered are those very rare ones where the high-energy electron has deposited most of its energy in the pixel of entry, causing a slight improvement in the MTF and DQE.
%This is similar to the behaviour observed in the Eiger HPD when using high-energy electrons \cite{Tinti2018}.

\section{The Influence of Defects on Sensor Performance}
\label{S:6}

\begin{figure*}[h!]
    \centering
    \includegraphics[scale=0.4]{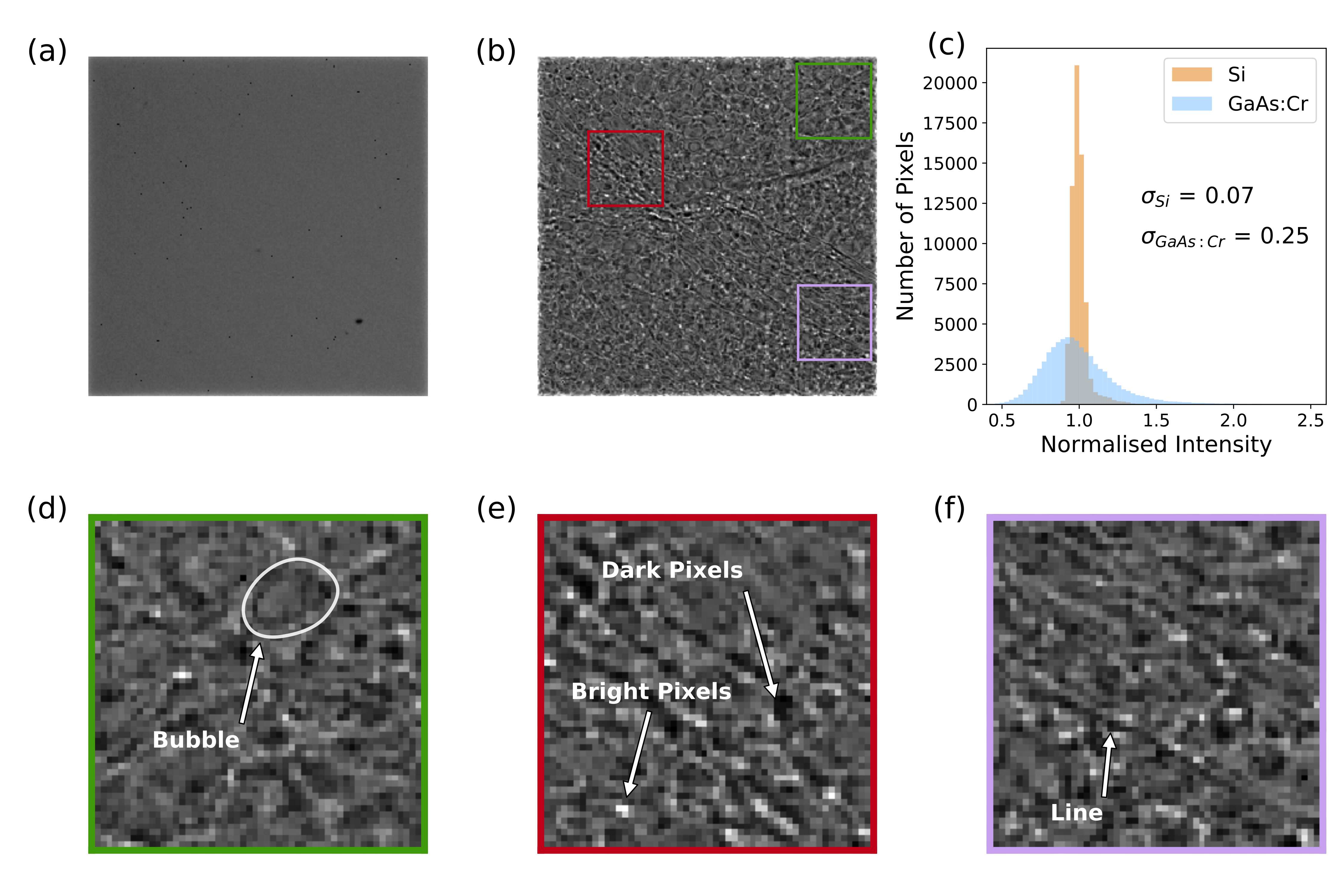}
    \caption{Flat field images of (a) Si and (b) GaAs:Cr devices operating in SPM with 200keV electrons normalised to their respective mean values; (c) histograms of the images in (a) and (b) with a note of the standard deviations of the intensity distributions; (d), (e), (f) show close-ups of regions of the GaAs:Cr sensor indicated in (b), with various types of defects in the GaAs:Cr sensor indicated. The contrast in all images has been adjusted such that the minimum and maximum intensities map to the limits of the x-axis in (c).}
    \label{fig:flat_fields}
\end{figure*}

\begin{figure*}[h!]
    \centering
    \includegraphics[scale=0.4]{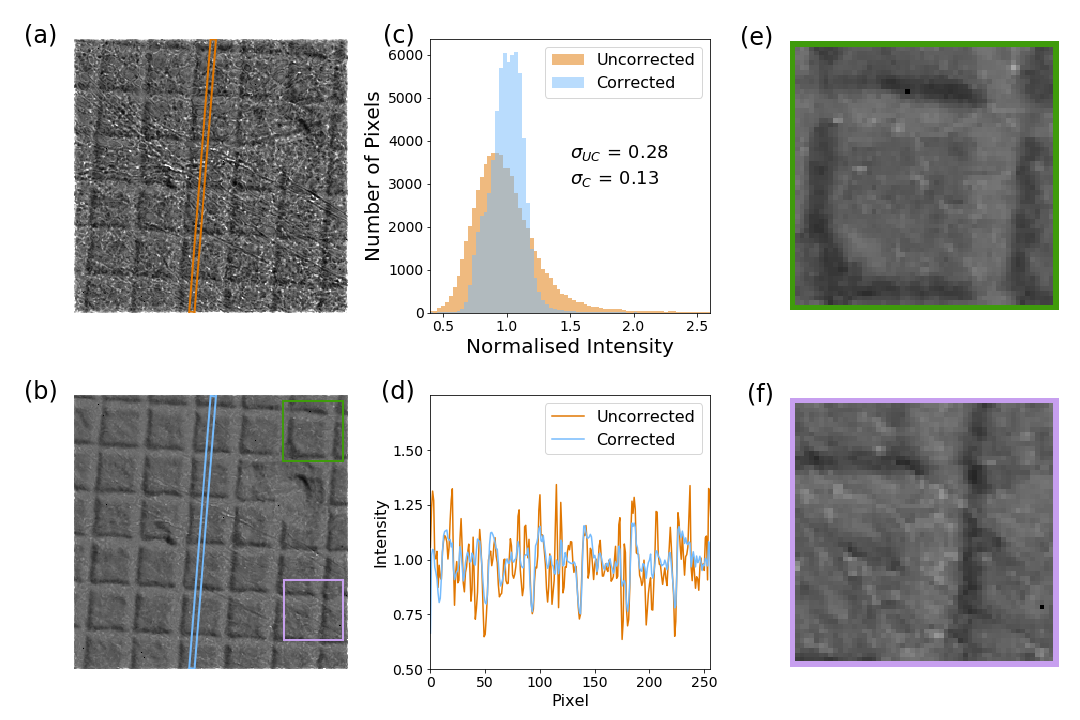}
    \caption{Low-magnification images of a carbon cross grating with Au shadowing recorded with the GaAs:Cr sensor operating in SPM using 300keV electrons (a) without a flat field correction applied and (b) with a correction applied, normalised by their respective mean value; (c) histograms of the images seen in (a) and (b), with the standard deviations of the normalised intensity distributions noted; (d) plots of the line profiles indicated by the region of interest marked in (a) and (b), the width of which is the integration width of the line profiles; (e) and (f) are close-ups of the regions of interest highlighted in (b). As in figure \ref{fig:flat_fields}, the minimum and maximum intensities in the images are also the x-axis limits for the histograms seen in (c).}
    \label{fig:cross_grating}
\end{figure*}

A difficulty in the fabrication of compound semiconductors are the defects that develop as part of the growth process \cite{Rudolph2005, Rotsch2005}.
Figure \ref{fig:flat_fields} shows normalised flat field exposures of Si and GaAs:Cr devices. 
The Si sensor is homogeneous whereas the GaAs:Cr sensor displays a high number of features that range from small bubble-like structures to lines that extend across the full sensor.
Examples of some of the structures that can be observed in the GaAs:Cr sensor are shown in figures \ref{fig:flat_fields}(d) and (f). 
These features are indicative of defects in the sensor which result in non-uniformities in the electric field across the sensor and consequently the pixel matrix, causing distortions in both the shape and the size of the pixels.
Areas of increased intensity indicate larger pixels, which count a disproportionate number of electrons while darker regions indicate pixels that are smaller and that have lost hits to neighbouring pixels.
An example of a cluster of pixels that are larger than expected and therefore overcount is seen in figure \ref{fig:flat_fields}(e).
It should be noted that pixels that were identified as noisy due to damage, failure in the manufacture process etc. were masked prior to the image being acquired, so this is not the cause of the bright pixels such as those seen in figure \ref{fig:flat_fields}(e).
Histograms of the intensities in the normalised flat field images in figure \ref{fig:flat_fields}(c) clearly show the broader range of intensities present in the GaAs:Cr flat field image compared with flat field image recorded by a Si device. 
The standard deviation of the range of intensities present in the GaAs:Cr flat field image is almost four times that of the flat field image recorded by the Si sensor in spite of the latter having a greater number of dead pixels. 
The variation in intensity in the Si flat field image is due to the slight variation in the threshold across the pixel matrix.

A standard procedure to correct for variation in intensity across a pixelated detector due to e.g. variation in the counting threshold across the pixel matrix is to apply a flat field, or gain, correction.
In figures \ref{fig:cross_grating}(a) and (b), images of a standard calibration sample recorded by a GaAs:Cr detector operating in SPM with and without a flat field correction are shown.
The effect of applying a flat field correction can be seen from the line profiles taken across the normalised images with and without the correction in figure \ref{fig:cross_grating}(d) as well as histograms of the images in figure \ref{fig:cross_grating}(c). 
Comparing the line profiles extracted from the corrected and uncorrected images there is an improvement in the contrast present in the corrected image.
In the case of the corrected image, the periodicity of the cross grating is apparent and can be readily measured as having a mean value of 43 pixels. 
This is not the case for the line profile extracted from the uncorrected image, where the additional variation in intensity makes the period of the cross grating difficult to discern.
The mean value of both line profiles is 1.0, but while the standard deviation of the corrected image line profile is 0.09, that of the uncorrected image line profile is 0.14, an increase of more than 50\%.
This reflects the effectiveness in the flat field correction at reducing noise due to the sensor defects.
Similarly, the range of intensities present in the corrected image is reduced, with the standard deviation of the range of intensities in the corrected image being less than half that of the uncorrected image.

However, examination of two of the regions highlighted in figures \ref{fig:flat_fields}(d) and (e), show that the distortions present in the uncorrected image can still be seen in the corrected image.
Increasing the counting threshold increased the visibility of the defects, which may be due to a reduction in blurring due to the improved detector PSF at higher counting thresholds.
Overall, although applying a flat field correction compensates for the variation in intensity due to the pixel size no longer being constant across the sensor, it does not correct for the geometric distortions due to variations in shape of the pixels caused by skewness of the sensor's electric field.
Consequently, artefacts are introduced and features in the images recorded by the GaAs:Cr device are distorted.

\section{Conclusions}
\label{S:7}

Our results confirm that high-Z sensors improve the performance of HPDs for high-energy electrons.
When operating in SPM and using a high threshold, the GaAs:Cr Medipix3 device that we have characterised is able to match and surpass the performance of an ideal detector in terms of its MTF for electrons with energies in the range of 60 - 200keV.
Using a high threshold has a negative effect on the device DQE, however, and by using a low threshold the GaAs:Cr SPM DQE for 200keV electrons is able to significantly exceed that of an ideal detector at high values of $\omega$, with its performance at $\omega_{N}$ surpassing the theoretical limit by the maximum extent possible.
This compares favourably with that the DQE of the Falcon 3 MAPS detector when using 200keV electrons \cite{Kuijper2015}, though it should be noted that $\omega_{N}$ for the monolithic device is higher than that of the Medipix3 due to its smaller pixels. 

Although the GaAs:Cr device outperforms the Si device for high-energy electrons, for electron energies below 120keV, its performance is poorer both in terms of MTF and DQE than the Si detector.  
The lower DQE can be attributed to increased backscatter, while the poorer MTF may be due to the distortions in the shape and size of the pixels dominating at low energies when the effective pixel size is very close to the physical pixel size even when using a low threshold.
Nevertheless, the difference in performance between Si and GaAs:Cr devices at low electron energies is modest, particularly comparing their MTFs when operating in SPM. 
Except for the most demanding of experimental conditions, there is likely to be minimal drawback in using the GaAs:Cr in experiments where it is necessary to use electrons with energies down to 60keV.
Our results also suggest that at lower energies there is the potential for a serious degradation in performance.
As such, there is unlikely to be any advantage in the use of high-Z sensors for scanning electron microscopy or for use in TEM imaging using electrons with energies lower than 60keV.
However, it may be possible to improve the efficiency of the GaAs:Cr device at low energies if the frontside contact were made of a lower-Z material than Ni, such as Al, but this would require advances in device manufacture.

At the high-energy range considered in this work, the performance of the GaAs:Cr shows a significant performance loss for 300keV electrons.
This motivates the investigation of other sensor materials with even higher values of Z, such as CdTe and CZT (average Z~=~50), though it is worth noting that using such materials will likely lead to further degradation in performance at low electron energies.
No single imaging detector will perform ideally across all the electron energies typically used in transmission electron microscopy, but on the basis of our results, HPDs with GaAs:Cr sensors may offer the best performance over the widest range of accelerating voltages presently used in TEM. 
Although the performance of the GaAs:Cr sensor at 300keV is poor, it is similar to that of the Si detector at 200keV, and we note that Medipix3 devices with Si sensors, as well as similar HPDs with Si sensors are routinely used at 200keV and 300keV in a number of applications \cite{Temple2018, Zhou2020}.
It is therefore likely that GaAs:Cr sensors can be used with 300keV electrons, offering improved MTF and DQE compared to a Si device used at 200keV and 300keV.

To summarise how HPDs with high-Z sensors can be best utilised, it is helpful to discuss the optimal detector settings.
It has been argued that for counting HPDs the optimum threshold is equal to half the primary electron energy so as to obtain the maximum enhancement of the MTF while negating the effects that using a high threshold has on the DQE.
However, we argue that the choice of threshold should depend upon the constraints imposed by individual experiments. 
In situations where spatial resolution is of paramount importance and where there are no dose constraints then a high threshold that optimises the MTF can be used. 
For example, a threshold of 101.7keV makes the GaAs:Cr detector an ideal detector in terms of its MTF operating in SPM for 200keV electrons, and the best MTF is obtained when using a threshold of 117.4keV. 
The value of MTF($\omega_{N}$) for these thresholds are 2.5 and 3.1 times that obtained with the lowest threshold (12.7keV) used. 
However, many of the experiments that DEDs enable are ones where dose is a constraint making detector efficiency key. 
In such cases the corresponding decrease in DQE($\omega_{N}$) to 19.7\% and 9.3\% of its maximum value when these thresholds are used is likely to be unacceptable, particularly when information at high $\omega$ is important.
%Even if high spatial frequency information is not a priority, the decrease in DQE(0) to 28.8\% and 7.6\% of its maximum value at these thresholds renders the

One class of experiments where high-Z sensors are likely to be particularly useful are time-resolved experiments. 
By reducing the lateral spread in the signal produced by the primary electrons, the temporal resolution of HPDs should be also be improved at high energies, as the scattering of incident electrons over multiple pixels means they can be counted in multiple frames when using short frames times \cite{Paterson2020}. 
Other experimental modes that would benefit from the use of HPDs with high-Z sensors include 4D-STEM modes that depend on precise measurement of the deflection of disks in the diffraction pattern due to either transmitted or Bragg-diffracted electrons. 
Precise measurements of such deflections can be achieved by template-matching \cite{Krajnak2016, Zeltmann2020}, and, in principal, this kind of analysis and the automation thereof would benefit from an improved detector PSF.
An enhanced PSF would also be beneficial for diffraction-based experiments where it is desirable to maximise the scattering angle subtended by the detector, and hence sampling of reciprocal space, by using a short camera-length while still clearly distinguishing between closely spaced diffraction spots or disks.

In addition to these experimental modalities, there is scope for HPDs with high-Z sensors for applications requiring high spatial resolution and large fields-of-view if the entry point of incident electrons can be localised in a way that does not have a detrimental effect on DQE, particularly if this can be done to sub-pixel accuracy.
With the exception of the Si detector for 60keV electrons, CSM fails to consistently identify the entry pixel for both devices, though there are cases for which using CSM could be regarded as advantageous.
For instance, when using 200keV electrons, the low threshold GaAs:Cr CSM MTF($\omega_{N}$) is 0.43.
The threshold at which the SPM MTF($\omega_{N}$) is equal to this is 62.7keV, at which DQE(0) is 0.68 and DQE($\omega_{N}$) is 0.18, while the low threshold CSM DQE(0) and DQE($\omega_{N}$) are 0.91 and 0.11.
Using CSM is therefore able to provide a comparable MTF and high-frequency DQE while enhancing the low-frequency DQE compared with using SPM.
Nevertheless, our results confirm the need for more sophisticated approaches to localise the entry point of the electron tailored to electrons with energies typically used in TEM.
%At lower energies, when it would be expected that the lateral spread of the signal due to incident electrons is more confined, the failure of CSM may be due the way in which timing information is used to identify the pixel of entry.
%In CSM, the pixel that registers the signal induced on it as going below threshold last is identified as having the most energy deposited on it and therefore, as the pixel of entry.
%This follows from the length of the signal pulse being proportional to the energy deposited on the pixel, but this neglects the fact that the scatter of the electron perpendicular to the pixel matrix means that 
The ability of DEDs to distinguish between detector noise and incident electrons has enabled the identification of the entry point of the electron to sub-pixel accuracy with monolithic devices  \cite{Kuijper2015, Mcmullan2009}, and research into how to achieve superresolution with other types of DEDs is an active field \cite{Ryll2019, Correa2020}.
Recent work with a Timepix3 detector with Si sensor has confirmed that it is possible to significantly enhance the MTF of HPDs for 200keV and 300keV electrons by identifying the entry pixel using a convolutional neural network \cite{VanSchayck2020}.
However, it is not clear whether reducing the scatter of incident electrons in the sensor by using high-Z sensors will facilitate localisation of the entry point to sub-pixel accuracy, or if the loss of information about the electron trajectory will make this task more difficult.

Our work clarifies the advantages of a GaAs:Cr sensor compared to traditional Si sensors and also highlights the effects of the distortions that defects in the GaAs:Cr sensor introduce into the images recorded.
Similar defects have been observed in other high-Z sensor materials, such as CdTe and CZT \cite{Pennicard2014, Maneuski2012, Veale2020}. 
To maximise the potential benefits of high-Z sensors these defects must be addressed, either by correcting these by post-processing or by further fabrication development.
Given the technical challenges improving the manufacture high-Z semiconductors suitable for radiation detection present, the former is a more viable solution in the short to mid-term.

\section*{Acknowledgements}
The authors gratefully acknowledge funding from the UK Science and Technology Facilities Council through the Industrial Cooperative Awards in Science \& Technology (CASE) studentship ``Next$^{2}$ Detection - Investigation of Hybrid Pixel Detectors for Future Transmission Electron Microscopy Imaging" (grant no. ST/P002471/1). 
The authors are grateful to Ms Nadia Bassiri, Mr Michael Perreur-Lloyd and Mr David Doak for their help developing hardware to mount detectors on microscopes. 
X.M. thanks Deutsche Forschungsgemeinschaft (DFG) for funding (grant no. MU 4276/1-1).
K.A.P. thanks Dr Anton Tyazhev and Ms Anastasia Lozinskaya of Tomsk State University for helpful discussions regarding the GaAs:Cr sensors. 
K.A.P. also thanks Dr Gary W. Paterson for helpful discussions regarding the calculation of the detectors' noise profiles.
The authors are grateful to Karlsruhe Nano Micro Facility (KNMF) for access to the FEI Titan 80~-~300 (S)TEM, and K.A.P and D.McG. acknowledge Quantum Detectors Ltd. for funding travel to KNFC as part of the Industrial CASE studentship that supports K.A.P. 
The authors also thank Quantum Detectors Ltd. for the loan of the Si Medipix3 device characterised as part this work. 
This work has been carried out within the framework of the Medipix3 collaboration.

%% References


\bibliographystyle{model1-num-names}
\bibliography{gaas_mpx3_paper.bib}






% \end{thebibliography}

\end{document}